\documentclass[10pt]{article} \usepackage{amsmath} \usepackage{amssymb}
\usepackage{epsfig}
\usepackage{textcomp}

\newcommand{\cH}{\mathcal{H}}

 at10pt

\begin{document}
\title {Quantum contextuality from a simple principle?} \author{Joe Henson\footnote{email: j.henson@imperial.ac.uk} } \maketitle
\begin{abstract}
In a recent article entitled ``A simple explanation of the quantum violation of a fundamental inequality,'' Cabello proposes a condition on a class of probabilistic models that, he claims, gives the same bound on contextuality for the KCBS inequality as quantum mechanics, and also rules out PR-box nonlocality.  He conjectures that the condition will also reproduce quantum limits on contextuality in other scenarios.  Here we show that the proposed principle is actually too weak to derive these results.  Cabello has implicitly assumed in the proofs that if all pairs in a set $C$ of events are pairwise exclusive (so that their probabilities must sum to less than 1), the set $C$ can itself be considered exclusive.  Perhaps surprisingly, this is not in general true in the general probabilistic framework under discussion.  With this extra assumption Cabello's proofs are sound.  Furthermore, it is established that CE holds in quantum mechanics, providing a reasonable and simple new principle that may characterise quantum non-contextuality in many scenarios.

\end{abstract}
\vskip 1cm

\section{Introduction}

In the past few years there has been much interest in formulating simple physical or information-theoretic principles that are obeyed by quantum mechanics (see \textit{e.g}~\cite{Hardy:2011}).  A characterisation of QM in terms of such principles rather than by the current, rather less intuitive, mathematical definitions of the theory, would provide an aid to understanding and, possibly, generalisation (for example to quantum gravity) and could also simplify the derivation of many results in quantum information theory.  One part of this task is the problem of characterising the set of probabilistic correlations between spacelike separated systems that are possible in quantum mechanics.  For example, while local hidden variable models bound the CHSH quantity (a measure of non-locality) to below 2 \cite{Clauser:1969}, QM bounds the quantity to $2\sqrt{2}$ \cite{Tsirelson:1980}, which is less than 4, the bound imposed by the no-signaling principle, saturated by the ``PR box'' \cite{Popescu:1994}.  Various candidate principles have been suggested to account for the apparent lack of PR box correlations in nature \cite{Dam:2005} \cite{Pawlowski:2009} \cite{Navascues:2010} \cite{Oppenheim:2010}.  In these studies, it is usual to formulate a general probabilistic framework that encompasses all possible no-signaling correlations (see \textit{e.g.}~\cite{Barrett:2007}).  Experiments are usually abstracted as a list of possible settings, each with a list of outcomes, for each separated system, sometimes picturesquely referred to as the settings of dials and states of lights on some instruments.

Another feature of QM, non-contextuality in Kocken-Specker (KS) \cite{Kochen:1967} type scenarios, has recently been treated in a similar way.  That is, as for non-locality, certain measures of non-contextuality go beyond the values allowed by ``non-contextual hidden variable'' (NCHV) models, but on the other hand QM limits these measures to less than some ``logical limit'' obeyed by all models in the most general conceivable class \cite{Klyachko:2008} \cite{Cabello:2010} \cite{Badziag:2011}\cite{Liang:2011}.  This leads to the question:  is there a simple physical principle that reproduces these quantum bounds on non-contextuality?

In a very recent article Cabello proposes an answer, in the principle ``Global Exclusive Disjunction'' or GE \cite{Cabello:2012}.  He claims to prove that the principle rules out a large and interesting class of trans-quantum non-contextual models (whilst the condition is clearly obeyed by QM).  This article is a brief response to Cabello's, and we refer the reader there for further details and references.  Below, a flaw in Cabello's proofs is pointed out.  However, the structure of the proofs still contains much insight that is correct: essentially the same proofs are valid if we assume another physical principle in place of GE.

\section{A generalised framework for non-contextual models}

First, let us note that the most general constraints on these non-contextual models are not based on such a compelling physical principle as no-signaling.   Let us start, as in the non-locality case, with an experiment represented by a number of experimental settings (``contexts'') and for each one a (possibly non-exhaustive) list of mutually exclusive outcomes to which we assign probabilities.  In the case of contextuality, we need more: the idea that some outcomes in different experimental contexts lie in a special relation to each other -- we think of them, in some sense, as instances of same event occurring in the two different contexts (note that `outcomes'' here includes sets of the most ``fine-grained'' outcomes).  In this case we make the minimal demand that they must have the same probability assignments.  This is true in quantum mechanics, for instance, for two outcomes represented by the same projection operator, which may nonetheless be present in two distinct lists of mutually exclusive alternatives for two different experimental contexts.

When we reduce to the description of an experiment to a set of dials and lights, however, the question of what outcomes should be considered ``the same'' in this sense becomes more opaque.  If for instance we have a system with outcomes $(a,b)$ and settings $(x,y)$, we could assume that the marginals $P(a=1|x,y)$ should equal $P(a=1|x,y')$ where $y \neq y'$.  With this ``compatibility'' assumption, we can derive constraints on the probabilities of the finest-grained events: here, for example, $P(a=1,b|x,y)+(a=0,b'|x,y') \leq 1$  \cite{Cabello:2012}.  But we must bear in mind that, in our rather abstract and general model, any such stipulation of compatibility is simply an additional assumption that is not based on anything deeper.  On the other hand, we \textit{can} compellingly ground such assumptions if we bring extra \textit{physical} assumptions into the problem: for instance via the no-signaling principle, by locating the outcome $a$ at a subsystem spacelike to setting $y$, in which case these two marginal probabilities must indeed be equal.

Nevertheless, \textit{given} some such compatibility assumptions we can make progress, and give a generalised probabilistic framework for KS-type non-contextuality scenarios. What is presented here is in substance the same framework as that of Cabello, Severini and Winter \cite{Cabello:2010}, but with some changes in notation and omission of irrelevant details (see \cite{Liang:2011} for another highly relevant approach to these issues).  Let us first define a list of what we will call \textit{events}, indexed by a set of indices $V=\{0,...n-1\}$ to which we are assigning probabilities.  These correspond to full specifications of all the outcomes and settings.  By assumption, some subsets of these events are \textit{exclusive}: as in the example above, only one of them can happen and so their the sum of their probabilities must be less than one.  Exclusive sets need not only be pairs in this general framework, and neither must they be derivable in some way from the pairs \cite{Cabello:2010}.  This turns out to be an important feature of this general framework, as we will see.

Call the set of these exclusive sets $\Theta$.  The pair $\Gamma=\{V,\Theta\}$ can be described is a \textit{hypergraph} \cite{Cabello:2010}, or more specifically an \textit{abstract simplicial complex}, because it clearly obeys the following defining condition: if $C \in \Theta$ then for all $D$ such that $D \subset C$, $D\in \Theta$ also.  Thus in the language of simplicial complexes, which will be adopted here, exclusive sets of two events are edges in $\Gamma$, those containing three events are triangles, and so on to tetrahedra, pentachora and beyond.

For our general probabilistic models of such situations, we want to assign probabilities to each event: there should be a map $P:V \longrightarrow [0,1]$.  But the exclusivity relations lead to a constraint on this map, that is, that $\sum_{i \in C}P(i) \leq 1$ $\forall C \in \Theta$.  This most general class of probability assignments consistent with our assumed relations of exclusivity is called E by Cabello \cite{Cabello:2012}.  Note that this is much less than a non-contextual ``hidden variables'' (NCHV) model demands: in that case, there must be a probability distribution over the set of all subsets of $V$ such that the probability of any subset containing an exclusive set is 0.  The probability of a given event is then the probability, in this distribution, that the given event is contained in that subset.  The reasoning behind this is that each event could be true or false in some hidden configuration which is ``merely revealed'' by the experiments, and the probability distribution ranges over the possible sets of all true events; then, only one in each set of exclusive events can possibly be true.  Following Cabello we call the class of models that obey this condition NCHV.  It is also the case that not every assignment of probabilities in the class E can be realised within QM, giving another class of allowed probability assignments for each such simplicial complex, as we will see.

Let us consider three important examples on five events, all of which are relevant to Cabello's arguments.  We will make use of the sum of the probabilities for the five nodes (that is, $V=\{0,1,2,3,4\}$) as a measure of contextuality:

\begin{equation}
\label{e:quant}
S=\sum_{i=0}^{4} P(i).
\end{equation}

\begin{figure}[ht]
\centering \resizebox{3in}{1in}{\includegraphics{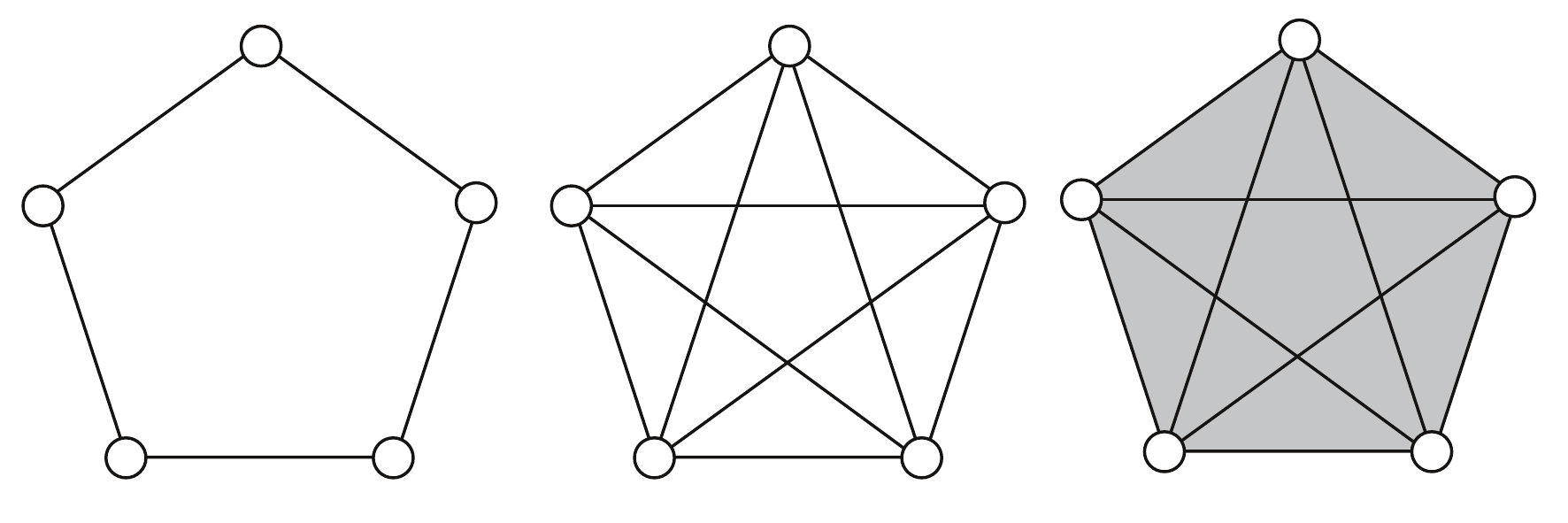}}
\caption{\small{
Three simplicial complexes on 5 vertices.  Vertices correspond to events and simplices to exclusive sets as explained in the main text.  On the left, there is the pentagon, or the cyclic graph on 5 vertices. The central diagram shows the pentagram, or the complete graph on 5 vertices.  The final diagram shows the pentachoron.  The shading is meant to indicate that, unlike the other two complexes, the edges are not the only simplices here:  all subsets of the vertices are simplices in the pentachoron, including the set of all five vertices.
\label{f:pent}}}
\end{figure}

The first example is the pentagon shown in figure \ref{f:pent}, the second is the pentagram shown next to it, and the third is the pentachoron.  For the pentagon, the exclusive sets are $\{i,i+1\}$ where addition is taken modulo 5.  The quantity $S$ given above, in this case, is bounded in the KCBS inequality \cite{Klyachko:2008}, which gives the maximum value for NCHV models as follows.  Examining the exclusivity relations, it is clear that only two events could be true at any one time (otherwise two of the true events would have to be connected by an edge).  Thus the maximum of the sum of the probabilities is 2 under the NCHV constraint.  In contrast the most general constraint E only demands that $P(i) + P(i+1) \leq 1$ $\forall i \in V$ where the addition on indices is taken modulo 5. This constraint is maximised by the assignment $P(i)=1/2$ $\forall i \in V$, and so the bound from E is 5/2.  It has been shown that the constraint from QM is $\sqrt{5}$ \cite{Klyachko:2008} \cite{Cabello:2010} \cite{Badziag:2011}.

Now let us consider the pentagram.  Here the exclusive sets are all the pairs of events.  NCHV models can only allow one event to be true with this complex, and thus the bound on \ref{e:quant} in this case is the minimal value, 1.  For E, however, \textit{the bound is still 5/2}.  This bound is again maximised by the assignment $P(i)=1/2$ $\forall i \in V$.  With this assignment it is clear that $P(i)+P(j) \leq 1$ $\forall i,j \in V$ and so this satisfies E.  So, while this scenario may seem very counter-intuitive, it breaks none of the rules of our general probabilistic framework.  

For the pentachoron, \textit{all} subsets of $V$ are exclusive sets, rather than just the pairs.  Now things are different.  The strongest relation imposed by E is that from the entire set, $\sum_{i=0}^{4} P(i) \leq 1$, immediately telling us the bound on $S$.  It is crucially important for the arguments below to note that the most general class of probability assignments, E, \textit{gives different bounds} on S for the pentagram and pentachoron.  The contrast between the results for the pentagram and pentachoron illustrates why, when dealing with the most general probabilistic framework, it is important to retain the idea of the exclusivity relation as a \textit{simplicial complex}, as in \cite{Cabello:2010}, and not just use the graph of exclusive pairs (the ``skeleton'' of $\Gamma$) as in Cabello's more recent article \cite{Cabello:2012}.  The richer structure of the simplicial complex has properties that cannot be derived from the graph alone, and this matters for the allowed probability assignments in the general case.

\section{Cabello's principle and proofs}

After this brief review, let us turn to the claims made in \cite{Cabello:2012}.  There, GE is defined as the condition that E should hold not only for the original events in the experiment under consideration but also for joint events of the original experiment and another experiment being carried out completely independently: that is, events for the joint set of outcomes and settings of the two experiments, under the condition that the probabilities are simply the product of some set of probabilities for each experiment separately.  To use the above framework for this situation, we must first ask, given the simplicial complexes of the two independent experiments, what is the relevant complex for the joint experiment?  The answer is an easy generalisation of Cabello's answer for the graph \cite{Cabello:2012}.  If a set of events for either of the two sub-experiments are exclusive then the joint events must be exclusive too.  This makes the complex the OR product of the two complexes.  This means that, if the two complexes are $\Gamma_1=\{V_1,\Theta_1\}$ and $\Gamma_2=\{V_2,\Theta_2\}$ then for the joint system's complex $\Gamma_{12}$, $V_{12}=  V_1 \times V_2$ and a set of $k$ joint vertices $\{\{i_1,j_1\},\{i_2,j_2\},...,\{i_k,j_k\}\}$ is in $\Theta_{12}$ if and only if $\{i_1,i_2,...,i_k\} \in \Theta_1$ or $\{j_1,j_2,...,j_k\} \in \Theta_2$.

We are immediately struck by a puzzle here.  As we have stated, E is merely a condition that makes sure that probability distributions defined on the events make sense -- it follows from the definition of probability distributions on the outcomes for each experimental context, and the assumptions about compatibility of measurements, as Cabello notes in his introductory discussion.  Now, if we have well-defined probabilities, obeying these rules, on two systems $X_1$ and $X_2$ and we take the product of them to form a joint probability distribution, then these joint probabilities must, trivially, also be well-defined in the same sense.  In other words they also obey E.  How, if E on the two separate systems implies E on the joint system, can demanding E for joint experiments (that is, GE) strengthen the constraints imposed by E?  To investigate this, let us examine one of the proposed proofs.

Cabello's Result 1 is that GE gives the quantum bound for the KCBS quantity, that is, eqn.(\ref{e:quant}) in the case of the pentagon graph.  To prove this he forms the OR product of two pentagon graphs.  This graph $\Gamma$ has 25 vertices corresponding to ``global events'', and Cabello shows that they can be divided into 5 non-overlapping subsets whose induced graphs are pentagrams.  Then Cabello states that ``[t]herefore, the sum of their [that is, the five events in one of the pentagrams] probabilities  cannot exceed 1''.  From this the sum of probabilities for all 25 event must be less than 5, ``and the only way to reach [this limit] is by assigning probability 1/5 to each and every one of the global events''.  Finally, since each of these events is the product of the probabilities from two identical experiments, the probability for the events in one of the separate sub-experiments must be $1/\sqrt{5}$, and so we have that $\sum_{i=0}^{4} P(i) = \sqrt{5}$, the quantum bound.  Result 2, that PR box non-locality \cite{Popescu:1994} is banned by GE, is a consequence of this proof, and so all the main results in the article hang on this reasoning.

The flaw in the proof is the statement that the sum of the probabilities of five events in a pentagram cannot exceed 1.  As we have already seen it is perfectly consistent in the general case to assign probability 1/2 to each event in a pentagram, giving a sum of $5/2$.  This resolves the earlier puzzle: GE is actually no stronger than E, and Cabello's claimed results are stronger than those which are actually proved. GE makes no restriction other than those made by probability theory applied to experimental outcomes (assuming the given compatibilities of measurements).  The problem came from using a graph instead of the richer concept of the simplicial complex to represent exclusivity relations, when considering the most general probabilistic models\footnote{Cabello offers another proof in the appendix which contains essentially the same flaw.  There, a result from \cite{Cabello:2010} is stated in a slightly incorrect form.  The ``fractional packing number'' $\alpha^*(G)$ of a \textit{graph} $G$ is, in the appendix of \cite{Cabello:2012}, defined as ``$\max \sum_{i\in V(G)} w_i$, where the maximum is taken over all $0 \leq w_i\leq 1$ and for all \textit{cliques} $C_j$ (subsets of pairwise linked vertices) of $G$, under the restriction $\sum_{i \in C_j} w_i \leq 1$''.  But in \cite{Cabello:2012} the packing number of a \textit{hypergraph} (or simplicial complex) was defined differently, with ``cliques'' in the definition replaced by what are here called simplices.  The effect is that results that should apply to pentachora are being stated for pentagrams, where they are not actually correct, in much the same way as in the primary proof.}.

\section{Consistent Exclusivity}

Nevertheless, Cabello's proof has much substance: it does provide a fascinating result, if one assumption is added.  \textit{This} assumption, implicit in Cabello's proof, is the condition that enforces the quantum limit, rather than GE.  The condition is here called \textit{Consistent Exclusivity}.  To define it, some auxiliary definitions are necessary.  A \textit{clique} is a set of vertices $C \subset V$ in which every pair is an edge in $\Gamma$.  The \textit{clique complex} of a simplicial complex $\Gamma$ is the complex with the same set of vertices, in which the simplices are exactly the cliques of $\Gamma$ (note that every simplex must be a clique by the definition of abstract simplicial complexes, and so the clique complex contains the original complex).

\paragraph{Consistent Exclusivity (CE):}  In a simplicial complex $\Gamma=\{V,\Theta\}$ representing the exclusive sets of events in an experiment, the assignments of probabilities to the events corresponding to $V$ obey E applied to the clique complex of $\Gamma$.

\paragraph{} Informally, if CE holds then ``we may as well have started with the clique complex of $\Gamma$ rather than $\Gamma$ itself when assigning probabilities''.  Or, again, if every pair of vertices in some set of vertices $C \subset V$  is exclusive, then if CE is true it will not affect the allowed probability assignments if we assume that the set $C$ is itself an exclusive set\footnote{Instead of stating the principle this way we could simply state that any simplicial complex that is not its own clique complex cannot represent the exclusive sets of events in an experiment.}.  It has previously been noted, in a similar context, that this principle (or something essentially equivalent to it) does not hold in the general case but does hold in QM \cite{Cabello:2012} (see also \cite{Liang:2011}).  The novel result here is that the principle turns out to be restrictive in an interesting way.

It is easy to see that assuming this condition completes Cabello's proof:  a pentagram subgraph in $\Gamma$ is a clique, and so in the clique complex of $\Gamma$, the induced complex on the same subset of vertices must be a 5-vertex simplex: a pentachoron.  In this case, Cabello's claim -- that the sum of the probabilities of five events in a pentagram cannot exceed 1 -- becomes true, as we have seen, and the rest of his proof follows.  Thus the counter-intuitive results for the pentagram which break Cabello's proof cannot affect the result when CE is assumed.  From this, using Cabello's Result 2, we can also see that CE bans PR boxes (or at least, any universe that contains more than one of them!).  See appendix \ref{a:pr} for further discussion of PR boxes and why GE alone cannot rule them out, and further comments on the relevance of the no-signalling constraint.

It remains to show that the probability assignments allowed by QM all obey CE.  This result is also very straightfoward.  In a quantum model of the type of experiment being considered here, the experimental outcomes for one experimental context correspond to projections onto orthogonal subspaces in a Hilbert space $\cH$.  That is, each set of outcomes maps to such a set of subspaces.  A set of events is exclusive if and only if the corresponding set of subspaces are all orthogonal to each other.  Then the result follows trivially: for a set of subspaces $C$ of a Hilbert space $\cH$, if each pair of subspaces represents an exclusive pair of measurements then each pair of subspaces in $C$ is orthogonal to every other.  In that case $C$ is a set of orthogonal subspaces, and thus can be added to the exclusive sets without further limiting the allowed probability assignments.

\section{Conclusion}

Although Cabello's GE is actually no stronger than the minimal requirement he calls E, adding CE to Cabello's conditions makes the proofs in \cite{Cabello:2012} sound and leaves us with an interesting result.  Here we have a condition on general experiments of the KS type, obeyed by quantum mechanics, which limits to the quantum bound for the KCBS quantity, and rules out PR boxes.  Both are important cases (see \cite{Cabello:2012} and references therein for more context).  Similarly to to the original claims, we might hope that the improved condition limits us to the quantum bounds for many other cases as well.  Cabello lists some interesting cases to check; it remains an open question how close CE comes to fully characterising quantum non-contextuality in such models.

Although QM always follows this rule (when the compatible measurements are defined in the usual way), in the general framework employed here (and which was used more explicitly in earlier work by Cabello \cite{Cabello:2010}) CE need not be true.  In particular, the system composed of two copies of the PR box does not obey this rule (when the compatible measurements are defined with reference only to the no-signaling constraint) even though it \textit{can} be consistently represented in the general probabilistic framework.  The point is that, for two copies of the PR box, the assumptions about what events must be exclusive that follow from these considerations of compatible measurements are counter-intuitively weak.

But what of the condition itself?  Is it in any sense a natural physical principle that we might expect to hold for any ``reasonable'' model?  Often, arguments for the naturalness or intuitiveness of such principles boil down to pointing out that the principle in question is true classically.  This is certainly the case here, in the sense that adding the higher-order simplices demanded by CE will clearly not alter the set of probability assignments allowed by NCHV for any graph.  Furthermore, CE seems like the least one would reasonably expect in terms of consistency between exclusive sets of events.  It is counter-intuitive enough that one cannot always consistently assign probabilities to the occurrence or non-occurrence of five events consistently with the exclusivity relations.  But to say that, even when every event in a set of $n$ events is mutually exclusive with every other, the sum of the probabilities of those events can \textit{still} be greater than 1, could justifiably be called a different order of counter-intuitive behaviour.  It is this rule that, for example, is violated by two copies of the PR box, as Cabello essentially showed in his article \cite{Cabello:2012}.  Like Information Causality \cite{Pawlowski:2009}, the CE principle is a remnant of classical physics that, it seems, still holds in the quantum world.  Hopefully, examining such principles will help us to understand the essential nature of that world.

\paragraph{} The author is grateful to Fay Dowker for pointing out the article to which this work is a response.  This work was made possible through the support of a grant from the John Templeton Foundation.

\section*{Note Added}

In the preprint \cite{Fritz:2012} (which appeared on the same day as that of Cabello \cite{Cabello:2012}, to which this article is a response), Fritz, Sainz, Augusiak, Brask, Chaves, Leverrier and Ac\'{\i}n (FSABCLA) define a principle dubbed ``local orthogonality'' (LO) which is, when the exclusive sets are defined with relevance to non-locality (as they are \textit{e.g.}~in the appendix of this article in the case of PR boxes), essentially the same as Consistent Exclusivity.  It is also demonstrated that QM obeys local orthogonality (similarly to results to be found in \cite{Cabello:2012} and \cite{Liang:2011} cited above).  Because of this previously overlooked precedence, and because \cite{Fritz:2012} contains and goes beyond the results given here on non-locality, the term local orthogonality should be preferred, although the word ``local'' is less apposite when applied to the other non-contextuality scenarios dealt with in Cabello's paper and, also, here.  This suggests the rather unwieldy ``compatible orthogonality'' as a compromise.  Viewed in this light, Cabello's work is, in a generalised context, implicitly making use of the same principle as FSABCLA.  A comment should also be made on Cabello's Observation 1 and its proof (ruling out two PR boxes).  While FSABCLA prove the observation directly by reference to an inequality that follows from LO (related by a relabeling to that referred to in figure \ref{f:pr} here), Cabello's proof first uses Result 1 to derive the KCBS inequality from the OR graph of two pentagons, and then finds such a pentagon in the graph for the PR box.  Cabello's result has thus divided the logic into two parts, the first being of more general application, rather than directly exhibiting an inequality for two copies of the PR box.  The results shown in the appendix here as a correction of Cabello's proof of Observation 2, where the two parts of the argument are brought back together for illustrative purposes, are therefore essentially the same as those previously shown by FSABCLA.  Because of this, it is no longer so clear that the observation, even in its two part form, should be considered a simplification of the results given \cite{Fritz:2012} as claimed directly above the statement of the observation.  In any case, the corrected version of Cabello's result 1 on the KCBS inequality is, at least to my knowledge, still new to the literature.

\bibliographystyle{plain}
\bibliography{refs}

\begin{thebibliography}{10}

\bibitem{Liang:2011}
Specker's parable of the overprotective seer: {A} road to contextuality,
  nonlocality and complementarity.
\newblock {\em Physics Reports}, 506(1-2):1--39, 2011.

\bibitem{Badziag:2011}
P.~Badziag, I~Bengtsson, A~Cabello, H~Granstrom, and J~Larsson.
\newblock Pentagrams and paradoxes.
\newblock {\em Found. Phys.}, 41:414, 2011.

\bibitem{Barrett:2007}
Jonathan Barrett.
\newblock Information processing in generalized probabilistic theories.
\newblock {\em Phys. Rev. A}, 75:032304, Mar 2007.

\bibitem{Cabello:2012}
A.~{Cabello}.
\newblock {A simple explanation of the quantum violation of a fundamental
  inequality}.
\newblock {arXiv:1210.2988}, 2012.

\bibitem{Cabello:2010}
A.~{Cabello}, S.~{Severini}, and A.~{Winter}.
\newblock {(Non-)Contextuality of Physical Theories as an Axiom}.
\newblock {arXiv:1010.2163}, 2010.

\bibitem{Clauser:1969}
John~F. Clauser, Michael~A. Horne, Abner Shimony, and Richard~A. Holt.
\newblock Proposed experiment to test local hidden-variable theories.
\newblock {\em Phys. Rev. Lett.}, 23:880--884, Oct 1969.

\bibitem{Fritz:2012}
T.~{Fritz}, A.~B. {Sainz}, R.~{Augusiak}, J.~B. {Brask}, R.~{Chaves},
  A.~{Leverrier}, and A.~{Ac{\'{\i}}n}.
\newblock {Local orthogonality: a multipartite principle for correlations}.
\newblock {arXiv:1210.3018}, 2012.

\bibitem{Hardy:2011}
L.~{Hardy}.
\newblock {Reformulating and Reconstructing Quantum Theory}.
\newblock {arXiv:1104.2066}, 2011.

\bibitem{Klyachko:2008}
A.~A. {Klyachko}, M.~A. {Can}, S.~{Binicio{\u g}lu}, and A.~S. {Shumovsky}.
\newblock {Simple Test for Hidden Variables in Spin-1 Systems}.
\newblock {\em Physical Review Letters}, 101(2):020403, July 2008.

\bibitem{Kochen:1967}
S.~Kochen and E.~P. Specker.
\newblock The problem of hidden variables in quantum mechanics.
\newblock {\em J. Math. Mech.}, 17:59, 1967.

\bibitem{Navascues:2010}
Miguel Navascués and Harald Wunderlich.
\newblock A glance beyond the quantum model.
\newblock {\em Proceedings of the Royal Society A: Mathematical, Physical and
  Engineering Science}, 466(2115):881--890, 2010.

\bibitem{Oppenheim:2010}
Jonathan Oppenheim and Stephanie Wehner.
\newblock The uncertainty principle determines the nonlocality of quantum
  mechanics.
\newblock {\em Science}, 330(6007):1072--1074, 2010.

\bibitem{Pawlowski:2009}
M.~{Paw{\l}owski}, T.~{Paterek}, D.~{Kaszlikowski}, V.~{Scarani}, A.~{Winter},
  and M.~{{\.Z}ukowski}.
\newblock {Information causality as a physical principle}.
\newblock {\em Nature}, 461:1101--1104, 2009.

\bibitem{Popescu:1994}
S.~Popescu and D.~Rohrlich.
\newblock Quantum nonlocality as an axiom.
\newblock {\em Found. Phys.}, 24:379, 1994.

\bibitem{Tsirelson:1980}
B.~Tsirel'son.
\newblock Quantum generalisations of bell's inequality.
\newblock {\em Lett. Math. Phys.}, 4:93--100, 1980.

\bibitem{Dam:2005}
Wim van Dam.
\newblock {Implausible Consequences of Superstrong Nonlocality}.
\newblock {arXiv:quant-ph/0501159}, 2005.

\end{thebibliography}

\begin{appendix}

\section{The PR box and Consistent Exclusivity}
\label{a:pr}

The PR box is describes a hypothetical experiment with two spacelike separated wings, in each of which there is a box with two settings, 0 and 1, and two possible outcomes for each setting.  The probability of finding outcomes $a,b$ given settings $x,y$ is 1/2 if $a \oplus b = xy$, where $\oplus$ is addition modulo 2, and 0 otherwise.

What are the relations of exclusivity in this case?  In this case, we have the ban on superluminal signaling as a basis for these relations.  That is, the event that $a=1$ given the setting in the same wing $x$ has the same probability whatever occurs in the other wing of the experiment, and we consider $a|x$ to be its own subsystem on this basis. This leads to a number of relations of exclusivity that must be obeyed.  Cabello picks out five events that lie in a pentagon of exclusive pairs for the PR box, which are all assigned probability 1/2.  It is on this basis that he claims the box is ruled out by GE.  Applying the earlier proof, the argument is that, for two copies of the PR box, we will be able to find a pentagram of ``joint events'' with probabilities that sum to more than 1.

\begin{figure}[ht]
\centering \resizebox{5in}{1.8in}{\includegraphics{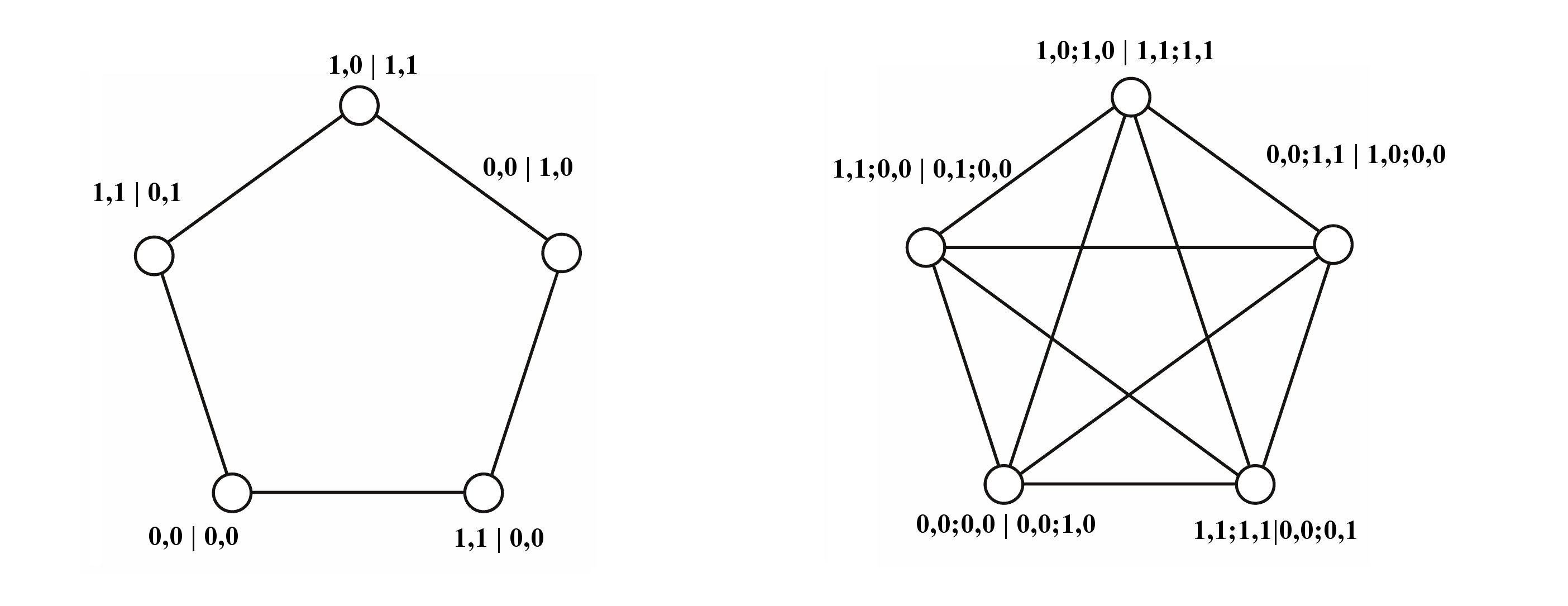}}
\caption{\small{
On the left, exclusivity relations for five particular events in a PR box (from the eight fine-grained events with non-zero probability) are illustrated.  On the right, five events from the joint system of two PR boxes are shown as vertices, with their exclusive sets as implied by the no-signaling constraint.  Here, in the notation $a,b;c,d|x,y,w,z$ the numbers given refer respectively to the two outcomes on the first and then the second PR box and, after the stroke, the settings on the first and second PR boxes respectively.  It is easy to check that every pair of these events is an exclusive set, and each one of them has probability 1/4.  Note that, in the language on the main text, this simplicial complex is a pentagram, not a pentachoron (that is, the set of all five events is not an exclusive set).
\label{f:pr}}}
\end{figure}

In figure \ref{f:pr}, this part of the argument has been made explicit, clearly illustrating the problem with the argument.  Here, five events from a joint system made up of two copies of the PR box are shown on the right.  Each event shown has probability 1/4 and every one is in an exclusive pair with every other.  Now, the question is, why should the probability of these five events sum to less than one?  With the set of five events made explicit, it is clear that the GE condition does not imply this.  No-signaling -- from which all exclusivity relations are derived in this case -- clearly does not imply the desired result: there is no signaling in a PR box on its own, and the probabilities in the joint system are just a product of the probabilities for the two copies.  In short, it does not follow from the general probabilistic framework under discussion that these probabilities should sum to one, and there is no physical principle invoked in the argument from which this follows.  It is necessary to assume the condition of Consistent Exclusivity in order to rule out this situation.

Note that it is easy to see from this diagram that QM rules out the PR box, as each of the five events illustrated would correspond to an orthogonal subspace in the Hilbert space of a quantum model.  That is, assuming these five events to be an exclusive set does not further limit the possible quantum models, but it does rule out the PR box probability assignments.

This illustrates an important problem for general models of non-contextuality, already touched on in the the main text: what justifies our assertion of the relations of exclusivity between events, if we are trying to conceive of a general operational framework?  For any experiment that we can model in QM, we can use the exclusive sets suggested by the quantum model and ask for the general class E of models given this graph.  But this is assuming something from QM.  If we are searching for a principle that constrains us from some general case back to QM, we do not want the answer to be made circular by sneaking theoretical structures from QM back into our general framework.  On the other hand, if we rely on the spacelike separation of subsystems to give us a sound physical basis on which to claim that some events must be exclusive, we find that this assumption on its own does not constrain from the general framework to QM.  Hence the task of discovering \textit{extra} reasonable principles which has become a major program in quantum foundations of late.  Nevertheless the condition of CE is not empty.  CE effectively strengthens the assumptions coming from no-signaling, by requiring that if there \textit{are} exclusive sets derived from no-signaling, then other exclusive sets must exist also.

\end{appendix}

\end{document}